\documentclass{article}
\usepackage[utf8]{inputenc}
\usepackage{amsmath}
\usepackage{amsthm}
\usepackage{hyperref}
\usepackage{amsfonts}
\usepackage[
    backend=bibtex,
    style=authoryear
]{biblatex}
\usepackage{wrapfig,graphicx}
\usepackage{braket}
\usepackage[dvipsnames]{xcolor}
\usepackage[margin=1in]{geometry}

\newcommand\citeposs[1]{\citeauthor{#1}'s\ (\citeyear{#1})}

\newcommand{\footremember}[2]{%
    \footnote{#2}
    \newcounter{#1}
    \setcounter{#1}{\value{footnote}}%
}

\title{Regularizing (away) vacuum energy}
\author{%
 Adam Koberinski\footremember{Waterloo}{Department of Philosophy, University of Waterloo, Waterloo, ON N2L 3G1, Canada}\footnote{atkoberi@uwaterloo.ca}  
  }
\date{Forthcoming in Foundations of Physics
}

\begin{document}
\newtheorem{theorem}{Theorem}[section]
\newtheorem{corollary}{Corollary}[theorem]
\newtheorem*{definition}{Definition}

\maketitle

\begin{abstract}
    In this paper I formulate Minimal Requirements for Candidate Predictions in quantum field theories, inspired by viewing the standard model as an effective field theory. I then survey standard effective field theory regularization procedures, to see if the vacuum expectation value of energy density ($\langle\rho\rangle$) is a quantity that meets these requirements. The verdict is negative, leading to the conclusion that $\langle\rho\rangle$ is not a physically significant quantity in the standard model. Rigorous extensions of flat space quantum field theory eliminate $\langle\rho\rangle$ from their conceptual framework, indicating that it lacks physical significance in the framework of quantum field theory more broadly. This result has consequences for problems in cosmology and quantum gravity, as it suggests that the correct solution to the cosmological constant problem involves a revision of the vacuum concept within quantum field theory.
\end{abstract}

\section{Introduction}
The cosmological constant problem has been a major focus of physicists working on theories of quantum gravity since at least the mid-1980s. The problem originates with unpublished remarks by Pauli, while interest in the problem increased in the 1980s due to inflation.  \textcite{Weinberg1989} famously laid out the the state of the field in the late 1980s, and used anthropic considerations to place bounds on the possible values of a cosmological constant in the Einstein field equations. The problem arises in a semiclassical merging of quantum field theory (QFT) and general relativity, where the stress-energy tensor for classical matter is replaced by an expectation value of the stress-energy tensor predicted by a particular model of QFT. When one does this, the vacuum expectation values of energy densities for each field have the same form as a cosmological constant term (i.e., a constant multiple of the metric), and so should contribute to the observed cosmological constant. However, when one takes a standard ``prediction'' of the combined vacuum energy densities from a model of QFT, the result is dozens of orders of magnitude larger than what is observed. Candidate solutions to the problem attempt to introduce new physics to reconcile the semiclassical prediction with observation; the predominant view in the physics literature is that an acceptable candidate for a theory of quantum gravity must solve the cosmological constant problem. Though many toy models have been proposed, there is no agreed upon solution pointing the way to the correct theory of quantum gravity.

The stubborn persistence of the cosmological constant problem provides motivation for a more detailed philosophical analysis of its assumptions. Assuming the ``old-fashioned'' view of renormalization, \textcite{KoberinskiCCP} breaks down the steps required to formulate the problem, and criticizes the justification behind each step. One of these steps involves the assumption that models of QFT predict the vacuum expectation value of energy density, $\langle\rho\rangle$. The prediction is taken to indicate that $\langle\rho\rangle$ is a physically significant quantity in the standard model. However, the problem changes shape when one accounts for the fact that the standard model is widely believed to be an effective field theory (EFT), with a built-in energy scale at which it breaks down. The EFT approach to QFTs makes sense of the old requirement of renormalizability, and uses the renormalization group equations to understand renormalization non-perturbatively.\footnote{For recent philosophical discussions of EFTs, see \textcite{WallaceQFT,WilliamsRG,JFraser2018,Ruetsche2020,KoberinskiSmeenkCCP}.}

As is well known, QFTs require renormalization in order to generate finite predictions. Renormalization consists of two steps: first, one introduces regulators to replace infinite quantities with quantities depending on an arbitrary parameter. The regulator $\mu$ must be such that (i) the regularized terms are rendered finite for all finite values of $\mu$, and (ii) the original divergent term is recovered in the limit $\mu \rightarrow \infty$. Next, one redefines some set of couplings such that the physically relevant value is independent of the regulator. Then the regulator is smoothly removed and the renormalized quantity remains finite. We say a model in QFT is renormalizable if all of its S-matrix elements can be made finite with a finite number of renormalized parameters. Even in a renormalizable model, vacuum energy density can only be regularized, but not fully renormalized. Since vacuum energy density is not a renormalizable quantity and plays no role in the empirical success of the standard model, \textcite{KoberinskiCCP} argued that one should not treat any value regulator-dependent value as a valid candidate prediction.

If, instead of predicting a value for $\langle\rho\rangle$, we simply expect the standard model to accommodate it as empirical input, the failure of naturalness prevents this weakened desideratum. In quantum electrodynamics (QED), for example, the electron mass and charge are renormalized to make the theory predictive. The theory takes these quantities as empirical inputs and therefore does not predict their values. Nevertheless, mass and charge are physically significant quantities in QED, necessary to the empirical success of the theory as a whole. Unfortunately, $\langle\rho\rangle$ cannot be input as an empirical parameter in the same way, due to its radiative instability order by order in perturbation theory. Further, since it plays no role in the empirical success of the standard model, there is little reason for $\langle\rho\rangle$ to play a central role analogous to mass and charge. Thus, if QFTs don't predict its value, it is best to understand vacuum energy density as outside their domain, and therefore not physically significant to QFT.\footnote{\textcite{KoberinskiSmeenkCCP} provide a more sustained argument that the cosmological constant problem signals a failure of naturalness for vacuum energy, in QFT and in general relativity as an EFT. The solution proposed there is to embrace new heuristics in theory construction, and to accept the limitations of the EFT framework for understanding fundamental physics.}

In light of the EFT view of the standard model, full renormalizability loses importance. If the standard model is an EFT, then (under the standard interpretation) it comes equipped with a physically significant cutoff scale and an infinite set of coupling constants consistent with the symmetries of the fields.The new couplings with mass dimension greater than four (in four dimensional spacetime) will be nonrenormalizable, but will have coupling constants that are suppressed by the momentum cutoff: $\alpha_i = g_i/\mu^n$. The explicit presence of the regulator in these terms is not a problem, since the regulator $\mu$ is much larger than the energy scales for which the effective theory is used. The renormalization group flow indicates that, at energies $E \ll \mu$, only the renormalizable terms have any appreciable effect. However, at higher energies, one may indeed see small deviations from the purely renormalizable terms, and these may be due to higher-order terms. Therefore, suitably regularized, nonrenormalizable terms can be physically significant when suppressed appropriately by a regulator.\footnote{Using precision tests of the standard model, one may find deviations from the predictions made using only the renormalizable terms. Examples of possible experimental tests include the anomalous magnetic moment of the electron or muon \parencite{Aoyama2012,KoberinskiSmeenkQED,Brookhaven} as well as the fine structure of positronium and muonium \parencite{PositroniumFineStructure}. In all of these cases, small deviations from the predictions made using the renormalizable standard model may be accounted for with higher-order couplings, suppressed by the physical cutoff scale.} Renormalizability is no longer a requirement, so long as the effects of nonrenormalizable terms become negligible at low energies.

If a suitably regularized vacuum energy density meets the requirements of a prediction in the EFT framework, then perhaps one is justified in claiming that the standard model predicts its value. There exist several regularization schemes for QFTs, and in general these will not agree on the algebraic form for any quantities until the renormalization procedure has been completed. Inspired by the EFT approach, and under the view that regulators are arbitrary, a suitable weakening of the requirement of renormalizability must satisfy the following requirements:\\

\textbf{Minimal Requirements for Candidate Predictions:} 
\textit{In order for a quantity within a model of QFT to count as a candidate prediction of some corresponding physical quantity, it must be the case that: (1) the quantity is largely insensitive to the regularization procedure; and (2) it is largely insensitive to changes to the value of the regulator.}\\

These requirements are motivated as follows. Violation of (1) would entail that different regularization schemes might be physically meaningful in that they encode different ways of parameterizing/forgetting high energy effects, and that for the quantity in question these differences matter. Supposing one views regularization schemes in this way, we learn that the quantity in question is sensitive to the physics at high-energies, and therefore does not fall within the proper scope of the EFT. Under the alternative view of regularization---as a formal tool used to render formally divergent terms finite---the independence of the predicted quantity from regularization scheme follows naturally. Under either interpretation, for an EFT to predict some quantity, it must satisfy (1).

Even though an EFT comes equipped with a physically significant cutoff energy scale, an important feature relevant to making predictions with EFTs is that the low-energy physics is largely insensitive to the exact value of that cutoff scale. In the context of the standard model, we are ignorant of the exact scale at which it breaks down. Any ``predictions'' from within the standard model that violate (2) are not true predictions at all; instead, they signify either that the quantity is meaningless when restricted to the low-energy EFT, or that it is highly sensitive to the details of the high-energy theory. In either case, one cannot say that the EFT predicts its value. Under the standard interpretation of EFTs, violation of (2) would signal that the EFT is insufficient to understand the phenomena in question. I will argue that the standard model $\langle\rho\rangle$ violates both minimal requirements, and this is best understood in the context of EFTs.

Physically significant quantities in a theory must be consistently described by that theory; if the standard model cannot provide a univocal, reasonable candidate prediction for the expectation value of vacuum energy density, then that failure is evidence that $\langle\rho\rangle$ is not physically significant in the standard model.\footnote{By physical significance of vacuum energy density, I mean the inference from a vacuum expectation value of an energy density term within a model of QFT to a real physical quantity onto which that value maps. One can believe that there is some real physical quantity of a suitably averaged value of vacuum energy density, to which our best physical theories don't accurately map (cf. \cite{SchneiderCCP}). The arguments in this paper undermine taking values from QFT to map onto the world; they say nothing about whether vacuum energy density exists. Undermining the physical significance of vacuum energy density for QFTs means that we should not trust that our best QFTs to accurately capture the relevant physics. Continuing the process discussed in \textcite{Saunders2002}, a further revision of the vacuum concept in QFT may be required, or perhaps even a full theory of quantum gravity.} Borrowing a common example of a classical fluid mechanics from \textcite{WallaceQFT}, we know that EFTs cannot predict all possible quantities relevant to the low-energy, macroscopic physics. In fluid mechanics, the formation of droplets and shock waves depend on the microphysical details of the fluid. We cannot use the effective theory of fluid mechanics to predict such behaviour, as the separability of scales breaks down. The underlying microphysical theory is then needed. Droplet formation and shock waves are physically real phenomena described by the microphysics, though fluid mechanics fails to describe them. I claim that the vacuum energy density $\langle\rho\rangle$ is a similar quantity that falls outside the domain of QFT. Vacuum energy may be a physically real phenomena, and some future theory may describe it, but it is beyond the scope of our best QFTs. The EFT framework helps to make this point more salient, because EFTs are explicitly meant to be limited in scope of applicability. The failure of $\langle\rho\rangle$ to satisfy either Minimal Requirement excludes it as a candidate for physical significance in QFT. Thus we should think of the cosmological constant problem as highlighting one limitation of our current best EFT. Since we are currently ignorant of the underlying microphysical theory to which the standard model is effective, there is little we can say about vacuum energy at present. In a separate paper \parencite{KoberinskiSmeenkCCP} I provide more general arguments that would lead one to a similar conclusion, and extends to the semiclassical merging of QFT and general relativity. My goal here is to show that, from within QFT as an EFT, $\langle\rho\rangle$ fails to meet the Minimal Requirements for a candidate prediction, and vacuum energy is therefore ill-defined until the future microphysical theory is known.

Though this conclusion is easiest to see within the EFT framework, the argument extends to QFT more broadly. \textcite{KoberinskiCCP} provides arguments for this conclusion in the context of the standard model as a fully renormalizable standalone QFT, and in Section~\ref{PointSplit} I argue that more rigorous extensions of QFT eliminate $\langle\rho\rangle$ from their conceptual framework, thereby supporting the conclusion that vacuum energy falls outside the domain of QFT, in any of its guises.

The strategy for the remainder of the paper is as follows. I provide a conceptual outline two major regularization and renormalization procedures that one might apply to extract a finite prediction of $\langle\rho\rangle$ from models of QFT, and discuss ways in which vacuum energy is removed in more rigorous local formulations of QFT. In Sec.~\ref{Orthodox} I consider the mainstream approaches to regularizing the standard model: lattice regularization and dimensional regularization. In Sec.~\ref{PointSplit} I consider some more mathematically rigorous approaches to QFT, and the ways that regularization and renormalization are treated there. In each case, I arrive at a value of $\langle\rho\rangle$ derived using that regularization scheme. Finally, in Sec.~\ref{Conclusion}, I compare the results to see if they satisfy the above requirements. As I will show below, purely regularized values of $\langle\rho\rangle$ satisfy neither Minimal Requirement, and we have no reason to accept a one-loop renormalized quantity as a candidate prediction either. Further, rigorous extensions of QFT that aim to provide a local description of fields remove the quantity $\langle\rho\rangle$ entirely, suggesting that vacuum energy falls outside the scope of QFT \textit{and} any merger of QFT and general relativity that emphasizes local covariance.

\section{Orthodox regularization of \texorpdfstring{$\langle \rho \rangle$}{<rho>}}
\label{Orthodox}
Standard cutoff regularization schemes in QFT require the inclusion of two momentum cutoffs: a lower bound to regulate the infrared divergences, and an upper bound to regulate the ultraviolet divergences. In position space, this is equivalent to defining the theory on a four-dimensional lattice in a box. Under the orthodox reading of EFT, the upper bound gains physical significance as the scale at which the effective theory breaks down.\footnote{The lower bound may be interpreted as encoding the fact that QFTs are only used in local regions of spacetime. Imposing some set of boundary conditions for long distances just means that we don't expect the model to apply in all of spacetime.} This view has recently been criticized \parencite{RosalerHarlander}, but is the dominant view of particle physicists and is becoming more mainstream amongst philosophers \parencite{Wallace2011,Williams,JFraser}. Below (Sec.~\ref{EFT}) I will outline the textbook approach to cutoff regularization in more detail, and discuss the modifications made to this formalism by the EFT view.

Historically, dimensional regularization was the favoured scheme for renormalizing Yang-Mills gauge models of QFT, like the electroweak model and quantum chromodynamics. Though it has received less philosophical attention due to its more formal nature, dimensional regularization is a powerful tool, and one that maintains Lorentz invariance. If one hopes to have a regularized candidate prediction of the vacuum energy density from the standard model, it should obey the correct equation of state that is required by the cosmological constant. Dimensional regularization gives this equation of state and Lorentz invariance, and the one-loop renormalized value $\langle\tilde{\rho}_{dim}\rangle$ (Eq.~\eqref{rhodimren}) calculated using dimensional regularization thus provides the best claim to a prediction of vacuum energy density from within the standard model. Thus, if any orthodox quantity serves as a candidate prediction for vacuum energy density, it is $\langle\tilde{\rho}_{dim}\rangle$. However, the instability of a one-loop renormalized vacuum energy density under radiative corrections indicates that naturalness fails here, and that vacuum energy may be sensitive to the details of high-energy physics.

\subsection{Momentum cutoffs and effective field theory}
\label{EFT}
For simplicity, I will illustrate the regularization techniques using a free scalar field theory, whose action is
\begin{equation}
    S[\phi,J] = -\int d^4 x \left( \frac{\eta^{\mu\nu}}{2}\partial_{\mu}\phi(x) \partial_{\nu}\phi(x) +\frac{m^2}{2}\phi^2(x) +J(x)\phi(x)\right),
\end{equation}
with $\eta_{\mu\nu}$ the Minkowski metric (here written with a $(-,+,+,+)$ signature), and the expression inside the integral is the Lagrangian density $\mathcal{L}$ for the model, plus source term $J(x)\phi(x)$. One can define a particular model of QFT with a built-in set of cutoffs, or one can impose cutoffs on individual expressions as the need arises. The former accords more closely with the EFT view, while the latter was standard in the early history of quantum electrodynamics, and remains standard in most introductory texts. Under the latter view, cutoffs are imposed in order to regulate divergences, and are removed from the renormalized theory.\footnote{For a more detailed analysis of the differences between the two approaches to renormalization, see \textcite{WilliamsRG,Rivat}. The latter argues that EFTs are best understood strictly under cutoff regularization. However, as I show below for the vacuum energy density, many features of QFTs are most easily understood under dimensional regularization.} We start with the latter approach to illustrate the algebraic form for expectation values of energy density and pressure.

In the case of calculating the energy density associated with the vacuum state, we are looking for the vacuum expectation value of the Hamiltonian density. In the case of the free scalar model, this is
\begin{equation}
    \langle \rho \rangle = \bra{0}\mathcal{H}\ket{0} = \frac{1}{2}\bra{0}\left( (\partial_t \phi)^2 + \delta^{ij}\partial_i\phi\partial_j\phi + m^2\phi^2 \right)\ket{0}.
\end{equation}
Using the Fourier expansion of $\phi$ one can calculate this to be (cf. \cite[IV.A][Eq. 68]{MartinCCPReview})
\begin{equation}
\label{rhoint}
    \langle \rho \rangle = \frac{1}{2(2\pi)^3} \int d^3 \mathbf{k} \omega_{\mathbf{k}},
\end{equation}
which diverges as $k^4$ for large $k$. Similarly, the pressure associated with the vacuum energy is
\begin{equation}
    \langle p \rangle = \frac{1}{6(2\pi)^3} \int d^3 \mathbf{k} \frac{k^2}{\omega_{\mathbf{k}}}.
\end{equation}
This is where one can regularize by introducing a momentum cutoff $\mu$, above which one no longer integrates. Doing so, one obtains the following expressions for the energy density and pressure:
\begin{align}
    \langle \rho \rangle &= \frac{\mu^4}{16\pi^2}\left[\sqrt{1+\frac{m^2}{\mu^2}}\left(1 + \frac{m^2}{2\mu^2} \right) - \frac{m^4}{2\mu^4}\ln\left( \frac{\mu}{m} +\frac{\mu}{m}\sqrt{1+\frac{m^2}{\mu^2}}\right) \right], \label{cutoffrho}\\
    \langle p \rangle &= \frac{\mu^4}{48\pi^2}\left[\sqrt{1+\frac{m^2}{\mu^2}}\left(1 - \frac{3m^2}{2\mu^2} \right) + \frac{3m^4}{2\mu^4}\ln\left( \frac{\mu}{m} +\frac{\mu}{m}\sqrt{1+\frac{m^2}{\mu^2}}\right) \right]. \label{cutoffp}
\end{align}

There are two things to note here. First, to leading order, both regularized terms depend on the cutoff scale to the fourth power. This regularization is therefore highly sensitive to what one takes as the cutoff scale, violating Reasonable Requirement (2). Under the old approach, one could renormalize $\langle\rho\rangle$ by introducing counterterms to remove any $\mu$-dependence. Unfortunately, the renormalized term does not carry over in a straightforward way to a field theory with interactions. Though one could simply define $\langle \rho_{physical} \rangle \equiv 0$ by subtracting off the entirety of the ``bare'' prediction, such a procedure is not stable against higher order quantum corrections. This holds true whether one subtracts off the entire prediction, or just the leading order divergent terms. In interacting theories, such as the scalar $\lambda\phi^4$ theory, the coupling between vacuum and gravity will contain contributions proportional to $\lambda$, $\lambda^2$, $\lambda^3$ and so on. If one defines $\langle \rho_{physical} \rangle$ to be independent of the cutoff scale at order $\lambda$, then equally large ($\sim \mu^4$) contributions spoil this cancellation at order $\lambda^2$, and so on for higher orders. So the value of $\langle \rho \rangle$ in Eq.~\eqref{cutoffrho} cannot be fully renormalized, and as it stands depends too sensitively on the (supposedly arbitrary) cutoff scale to count as a prediction.

Second, notice that the ratio $\langle p \rangle /\langle \rho \rangle \neq -1$, as one would expect from a Lorentz-invariant vacuum. This is because the cutoff procedure is itself not Lorentz-invariant. In order to obtain a vacuum energy density that respects the Lorentz symmetry and reproduces the equation of state required by a cosmological constant term, one must subtract the leading order $\mu^4$ terms in each, which is only justified in the context of modified minimal subtraction schemes using dimensional regularization.

The above discussion is framed in the old-fashioned context of ad-hoc regularization. What changes when we think of QFTs as EFTs, where the cutoff plays a more direct role? In the EFT framework, a QFT is defined with a built-in UV cutoff. To make the overall theory finite, an IR regulator is often used, though this may be smoothly removed at the end of the calculation to return to a continuum theory. I start with both regulators, which effectively places the field theory on a Euclidean lattice, converting the integrals in the action and the over field operations into discrete sums. For 4D lattice spacing $a$, placing the model in hypercube of length $L$, the generating functional becomes
\begin{align}
    \mathcal{Z}[J] &= \displaystyle \int_{\mu} \mathcal{D}\phi \exp\left( i \int d^4 x [\mathcal{L}(\phi(x)) + J(x)\phi(x)]\right) \\
    &\equiv \int \displaystyle\prod_{l=1}^{N}d\phi_l \: \exp\left(ia^4 \sum_{l=1}^{N}[\mathcal{L}(\phi_j) + J_l \phi_l] \right),
\end{align}
where $N= (L/a)^4$ and $\mu = 2\pi/a$. The quantities $a$ and $L$ are built-in ultraviolet and infrared regulators. Once a set of fields is specified, along with the expected symmetries of the model, the Lagrangian is defined to include all terms involving the chosen fields and respecting the symmetries; this means that the Lagrangian is likely to be a formally infinite sum of terms, each multiplied by its own coupling constant. As initially stated, this would be a major problem; though the path integral has been IR and UV regulated, we now have an infinite number of terms in the Lagrangian. There is no a priori reason to expect that the bare coupling parameters decrease for higher-order field contributions, and thus no indication of an appropriate truncation of terms in the Lagrangian.

However, one uses the renormalization group transformations to rewrite the generating functional in terms of a new, lower ultraviolet cutoff $\mu'=\mu-\delta\mu$. One separates the integral over field configurations $\int_{\mu}\mathcal{D}\phi \rightarrow \int_{\mu'}\mathcal{D}\phi_{\mu'} \int_{\delta\mu}\mathcal{D}\phi_{\delta\mu}$, and integrates out the field modes $\phi_{\delta\mu}$. The amazing feature of the renormalization group is that, when one does this, the new expression for the Lagrangian retains the same form. All of the effects of the field modes above the new cutoff can be absorbed into a redefinition of the coupling constants in the Lagrangian. Since coupling constants will be dimensionful quantities (the Lagrangian has units of $[\mathrm{energy}^4]$, and scalar fields have dimensions of energy) redefinitions of coupling involve powers of the new cutoff scale. If the cutoff scale is large compared to energy levels of interest for the effective theory, then higher-order terms in the Lagrangian will be suppressed by the new coupling constants $g_i \rightarrow g_i/(\mu')^n$. In the limit where energy scales of interest are vanishingly small compared to the cutoff, all terms with high powers of fields and their derivatives will be suppressed by inverse powers of the cutoff.

Though there is much more to be said about the renormalization group and EFT, there are two major points relevant to the discussion of regularizing vacuum energy. First, one defines a model in EFT with built-in regulators. Renormalization is no longer a primary focus, since the renormalization group techniques indicate the irrelevance of most nonrenormalizable terms. Since regulators are present in the definition of the theory, one needn't worry about regulators appearing in predictions. As long as the predictions do not depend sensitively on the precise value of the cutoff---since the value of the physically meaningful cutoff is unknown until a future successor theory is developed---its presence is not a problem in the EFT framework. Thus, the EFT framework motivates Minimal Requirement (2) discussed in the Introduction. However, the vacuum energy is still a problem, since it depends sensitively on the cutoff---as mentioned above, $\langle \rho \rangle \sim \mu^4$. The problem of renormalization changes dramatically under the EFT view, since the presence of $\mu$ in Eq.~\eqref{cutoffrho} is not in itself a problem. The momentum cutoff is standardly taken to have physical significance for the future successor theory; there is therefore no reason to renormalize by subtracting the $\mu^4$ term, and so even an illusory insensitivity is to $\mu$ is lost.

Second, by defining models of QFT with a built in lattice scale, issues of Lorentz invariance may lose importance. If the lattice is to be physically significant, then Lorentz invariance of EFTs only holds approximately. Accordingly, one would not expect the vacuum energy density to be exactly Lorentz invariant, and so the concern regarding the wrong equation of state from Eqs.~\eqref{cutoffrho} and~\eqref{cutoffp} is less pressing. However, the failure of exact Lorentz invariance would undermine the motivation to subtract off only the $\mu^4$ term for a one-loop renormalization, and it would be much harder to input the vacuum energy density into the Einstein field equations. If straightforwardly input into the Einstein field equations as is, one would get an entirely different equation of state for the cosmological constant. Given that the EFT framework is predicated on the idea that physics at disparate energy scales separates, it would be curious if a consequence of that framework was that small scale violations of Lorentz invariance implied qualitative changes to physics on cosmological scales. In any case, failure of Lorentz invariance would undermine the standard motivations for the cosmological constant problem, though the presence of an enormous vacuum energy density for the standard model would remain.\footnote{The fact that Lorentz invariance is lost if the lattice structure of effective field theories is taken literally should have observable consequences. Incredibly sensitive tests have failed to detect violation of Lorentz invariance at small scales \parencite{Mattingly}. Though outside the scope of this paper, one might argue that a literal interpretation of the lattice is therefore unmotivated from the point of view of both QFTs and general relativity.}

\subsection{Dimensional regularization}
\label{DimReg}
Dimensional regularization has historically played an important role in the development of the standard model. \textcite{tHooftVeltman} first proved that Yang-Mills gauge models are renormalizable by developing and employing dimensional regularization. The method is often more powerful, since the symmetries of a model---both gauge symmetries and spacetime symmetries---remain intact. It allows for an easier identification of divergences than the momentum cutoff approach, and naturally suggests a minimal subtraction (or, alternatively, modified minimal subtraction) method of renormalization. Finally, this method also removes infrared divergences associated with massless fields without introducing a further regulator. The disadvantage is that a physical interpretation for the regulator is rather opaque; the method is more clearly formal than the momentum cutoff approach.\footnote{This is only a disadvantage if one expects a regulator to be physically significant. If regularization is treated simply as a procedure for taming divergences, then the regulators need not have a physical significance. Further, if the analogy between lattice regularization in condensed matter physics and particle physics is misleading, then the physical interpretation that lattice regularization provides may actually lead to an unjustified physical interpretation (cf. \textcite{FraserRG,FraserKoberinski}).}

In the case of the vacuum energy density one aims to include its expectation value in the Einstein field equations. It is therefore important to ensure that the Lorentz symmetry of the expression is maintained---since it is this feature of $\langle\rho\rangle$ that justifies its interpretation as a contribution to the cosmological constant. Dimensional regularization is best suited for this purpose. I will outline the regularization technique for vacuum energy for a scalar field. As \textcite[Sec. VII]{MartinCCPReview} demonstrates, the calculations for fermions and gauge bosons proceeds in a similar fashion, though the leading multiplicative coefficients (of $\mathcal{O}(1)$) differ.

The integral for energy density in Eq.~\eqref{rhoint}, in $D$-dimensional spacetime becomes
\begin{align}
    \langle \rho \rangle &= \frac{\mu^{4-D}}{2(2\pi)^{D-1}}\int d^{D-1}\mathbf{k} \: \omega_{\mathbf{k}} \\
    &= \frac{\mu^{4-D}}{2(2\pi)^{D-1}} \displaystyle\int_{0}^{\infty} dk \: d^{D-2}\Omega \: k^{D-2}\omega_{\mathbf{k}},
\end{align}
where $d^{D-2}\Omega$ is the volume element of the $(D-2)$-sphere, and the $\mu$ is an arbitrary scale factor such that the equation has the right unit dimensions.\footnote{I use $\mu$ as an arbitrary scale factor here because it appears in the formal expression for $\langle\tilde{\rho}_{dim}\rangle$ in the same way that the (arbitrary) momentum cutoff appears in the lattice regularized expression. The fact that these scales have different meanings supports my argument that these terms differ significantly. The same term for the regulator is used simply to aid algebraic comparison.} Using the fact that the general solution of angular integrals can be expressed in terms of gamma functions, the solution to this integral is
\begin{equation}
    \langle \rho \rangle = \frac{\mu^4}{2(2\pi)^{(D-1)/2}} \frac{\Gamma(-D/2)}{\Gamma(-1/2)} \left( \frac{m}{\mu}\right)^D.
\end{equation}
Performing the same operation for the pressure, one obtains
\begin{align}
    \langle p \rangle &= \frac{\mu^{(4-D)}}{2(D-1)(2\pi)^{D-1}} \int d^{D-1}k \: \frac{k^2}{\omega_{\mathbf{k}}}\\ 
    &= \frac{\mu^4}{4(2\pi)^{(D-1)/2}} \frac{\Gamma(-D/2)}{\Gamma(1/2)} \left( \frac{m}{\mu}\right)^D.
\end{align}
Since $\Gamma(-1/2) = -2 \Gamma(1/2)$, we obtain the correct equation of state, $\langle p \rangle / \langle \rho \rangle = -1$. If one expands the gamma functions in the above expressions, and sets $D = 4 - \epsilon$, then the regularized $\langle\rho\rangle$ and a one-loop renormalized expression $\langle\tilde{\rho}_{dim}\rangle$ are
\begin{align}
    \langle \rho \rangle &\approx -\frac{m^4}{64 \pi^2}\left(\frac{2}{\epsilon} +\frac{3}{2} - \gamma - \ln\left[ \frac{m^2}{4\pi\mu^2}\right] \right) + \cdots \label{rhodimreg} \\
    \langle \tilde{\rho}_{dim} \rangle &= \frac{m^4}{64\pi^2}\ln\left(\frac{m^2}{\mu^2}\right), \label{rhodimren}
\end{align}
where $\gamma \approx -0.57772$ is the Euler-Mascheroni constant (cf. \cite[IV.A]{MartinCCPReview}, renormalized using modified minimal subtraction).\footnote{This is a first-order renormalized calculation. As \textcite[Sec. VI]{MartinCCPReview} highlights, this prediction is largely unchanged under a Gaussian approximation to an interaction term (i.e., to one loop). Since the expression remains the same, I refer to Eq.~\eqref{rhodimren} as a one-loop renormalized term.} This expression actually agrees (up to constants of $\mathcal{O}(1)$) with the leading order logarithmic term predicted using the momentum cutoff approach in Eq.~\ref{cutoffrho}, after subtraction of the $\mu^4$ term. \textcite{MartinCCPReview} notes that ``it is well-known that the dimensional regularization scheme removes the power law terms,'' (p. 13) so this is not a surprising result. Like in the case of Yang-Mills gauge models, dimensional regularization leaves the underlying symmetries of the model intact, and leads to a correct regularization that respects those symmetries. We see that, instead of a functional dependence on the fourth power of the cutoff, the vacuum energy density for a given field depends on the fourth power of the \textit{mass} of that field. This means that massless fields (photons, gluons) do not contribute to the dimensionally regularized or renormalized vacuum energy, at least to leading order.

It turns out that fermion fields and boson fields share this functional dependence, though each contains a numerical factor $n_i$ to multiply $\langle \rho_{ren} \rangle$. For the Higgs scalar, $n_H = 1$; for fermions, $n_F = -4$; for bosons, $n_B = 3$. \textcite[IX][Eq. (516)]{MartinCCPReview} determines the vacuum energy density coming from vacuum fluctuations (ignoring early universe phase transitions) to be
\begin{equation}
\label{SMrho}
   \langle \rho_{SM} \rangle = \sum \langle \tilde{\rho}_{dim} \rangle = -2 \times 10^8 GeV^4,
\end{equation}
assuming a scale factor $\mu \approx 3 \times 10^{-25} GeV$, though the prediction is relatively insensitive to the exact value of $\mu$. This therefore seems like an impressive renormalization and prediction of the vacuum energy from the standard model. Since modified minimal subtraction is a natural procedure for dimensional regularization, the renormalization method is also justified. However, this term is renormalized to one loop; radiative instability will spoil renormalization at higher orders, and thus naturalness fails here as it does for lattice regularization. In general, the contributions from next-to-leading order for $\langle \tilde{\rho}_{dim} \rangle$ will be large enough to spoil the renormalization performed at leading order. The functional form of of Eq~\eqref{rhodimren} hides the high sensitivity to the regulator that appears at higher orders.

If we treat the standard model as an EFT, we may be justified in trusting predictions of some quantities only up to one-loop. As an example, the Fermi theory of weak interactions in now known to be an effective approximation to the electroweak model, valid for energies far less than the mass of W and Z bosons.\footnote{This example is discussed in more detail in Sec.~\ref{Conclusion}.} The Fermi theory is well-behaved up to one-loop, but is nonrenormalizable and badly divergent beyond this scale. The difference with the vacuum energy density is that $\langle \rho \rangle$ displays the same types of nonrenormalizable divergence at every order, while more severe divergences occur in the Fermi theory only at higher order than the one-loop terms.

The proper focus of our attention should therefore be the regularized term (Eq.~\eqref{rhodimreg}). As should be obvious by inspection, this value displays a sensitive dependence on the regulator $\epsilon$, and differs markedly from the lattice regularized quantity (Eq.~\eqref{cutoffrho}). Thus $\langle\rho\rangle$ fails to satisfy either Minimal Requirement under orthodox approaches. One might argue that this failure is worse in the EFT framework, since EFTs are explicitly constructed to exclude contributions from certain energy scales. In the next section, I use more rigorous extensions of standard QFT to show that, even outside of the EFT framework, one should not expect QFTs to describe vacuum energy.

\section{Splitting hairs: splitting points}
\label{PointSplit}
Outside of the mainstream work in QFT and particle physics, there has been persistent effort to place the QFT formalism on more secure mathematical footing. One major goal of this work is to be clear about the validity of assumptions and algebraic manipulations standardly employed in particle physics. Point-splitting procedures are used to track more carefully the ways in which quantum fields---as operator-valued distributions---are multiplied together at coincident points. The project of doing QFT on curved spacetimes likewise demands a re-examination of the assumptions that go into constructing QFTs in Minkowski spacetime. In this section I discuss the Epstein-Glaser point-splitting procedure as a candidate regularization scheme, and consider the modifications needed to put QFT on curved spacetimes, a project largely pursued by Hollands and Wald. The modifications necessary indicate that Minkowski spacetime is particularly special, and that significant alterations to QFT may be needed even for a semiclassical merging with general relativity. If one hopes for an extension of QFT beyond the EFT framework, approaches like these are a likely first step. We see in both approaches that the vacuum energy concept does not arise, indicating that $\langle\rho\rangle$ is not a meaningful concept in QFT as a whole.

\subsection{Minkowski background}
Point splitting and other local approaches to regularization stem from \citeposs{Wilson69} early work on the operator product expansion, which is a formalism for defining products of operator-valued distributions at coincident points. Since we are concerned here with short distance behaviour of fields, the work in this tradition uses the position space representation of quantum fields. In ordinary approaches to QFT, distributions are not carefully handled, and this leads to divergences in products of operators at the same point. Wilson originally proposed an ansatz that two operators $A$ and $B$ defined at coincident points should be described by
\begin{equation}
    A(x)B(x) = \displaystyle\lim_{\chi \rightarrow 0}A(x+\chi/2)B(x-\chi/2) = \lim_{\chi \rightarrow 0}\sum_{i=1}^{n} c_i(\chi,x)C_i(x) + D(x,\chi),
\end{equation}
with $C_i(x)$, $D(x,\chi)$ local operators without divergences, and $c_i(\chi,x)$ coefficients that diverge in the limit $\chi \rightarrow 0$. The original operator product is then replaced with the regularized product
\begin{equation}
    \left[ A(x+\chi/2)B(x+\chi/2) - \displaystyle\sum_{i=1}^n c_i(\chi,x)C_i(x) \right]/c_n(\chi,x),
\end{equation}
which goes to zero as $\chi$ goes to zero.

Further work on the general properties of products of distributions---as mathematical physicists came to understand that quantum fields are operator-valued distributions---led to the Epstein-Glaser approach to regularizing and renormalizing QFTs. The conceptual move here involves switching focus from products of observables in neighbouring points to the products of fields at coincident points.

\textcite{EpsteinGlaser} proved---through more careful analysis of the properties of the S-matrix---that a renormalized perturbation theory could still obey microcausality and unitarity. Though a more mathematically technical and indirect regularization method, this approach tames many UV divergences present in QFT, and therefore accomplishes renormalization in a similar way. Essentially the n-point functions must be appropriately smeared with test functions $f(x_1, \ldots , x_n) \equiv f(x)$. Infrared divergences are dealt with by carefully removing the test functions in observable quantities; one takes the adiabatic limit $f(x) \rightarrow 1$ after constructing appropriate integrals.

Instead of treating point-splitting as a more mathematically elegant form of renormalization, \textcite[3.1,3.2]{Scharf} takes the causality condition for distributions to point to the correct method for defining the n-point distributions $T_n(x_1,\ldots , x_n)$ when the set of $\{T_m | \; 1 \leq m \leq n-1\}$ are known.\footnote{The treatment of point-splitting in this section follows the presentation in \textcite[Ch. 3]{Scharf}.} These n-point distributions are related to the perturbative construction of the S-matrix as follows:
\begin{equation}
    S(f) = \mathbf{1} +\displaystyle \sum_{n=1}^{\infty} \frac{1}{n!}\int d^4 x_1 \cdots d^4x_n T_n(x_1, \ldots, x_n) f(x_1) \cdots f(x_n),
\end{equation}
where $f$ is a complex-valued test function, and where the limit $f \rightarrow 1$ is taken at the end of the calculation. The causality condition is applied to the test functions as follows. Suppose there exists a reference frame in which $f_1$ and $f_2$ have disjoint supports in time,
\begin{equation}
    supp \: f_1 \subset \{x \in \mathbb{M}\: | \: x^0 \in (-\infty, r) \} \quad supp \: f_2 \subset \{x \in \mathbb{M}\: | \: x^0 \in (r, \infty) \}.
\end{equation}
Then the causality condition is the requirement that $S(f_1 + f_2) = S(f_2)S(f_1)$.

The $T_n(x_1,\ldots ,x_n)$---operator-valued distributions---are constructed by induction. One simplifies the procedure by decomposing the $T_n$ into (normal-ordered) free fields and complex number-valued distributions

\begin{equation}
    \label{Eq:DistFields}
    T_n(x_1, \ldots , x_n) = \displaystyle\sum_k : \prod_j \bar{\psi}(x_j) t_n^k(x_1, \ldots , x_n) \prod_l \psi(x_l) : : \prod_m A(x_m) : ,
\end{equation}
where $t_n^k$ is the momentum space numerical distribution. Now the problem switches from defining an appropriate splitting procedure for the $T_n$, to the simpler problem of defining a splitting procedure for the $t_n^k$. The usual procedure---in standard versions of interacting QFT---involves splitting with a series of $\Theta$ functions for each $x_i \in \{x_n\}$, but this is discontinuous as $x_i = 0$. If $t_n^k$ is singular for some $x_i = 0$, then the product is not well defined, and UV divergences appear. Instead, one introduces the concept of a scaling dimension $\omega$, signalling the degree of divergence for the distribution. This scaling dimension carries over to momentum space representations as well.

For QED in momentum space, distributions properly split have a series of free parameters, being defined only up to a polynomial of rank $\omega$.\footnote{It is possible that $\omega$ will not be an integer for some distributions, though this does not occur in QED. When $\omega$ is not an integer, the polynomial will be rank $\omega'$, the largest integer that is less than $\omega$.} The ``regularized'' distributions therefore take the form
\begin{equation}
    t(p) = t'(p) + \displaystyle\sum_{|a| = 0}^{\omega} C_a p^a
\end{equation}
after splitting, where $t'(p)$ is defined by the causality condition. The free parameters $\{C_a\}$ can be fixed by an appropriate choice of regulator on the distribution, and this is why, for all practical purposes, the causality condition is a more mathematically rigorous way to introduce regulators into the theory. Though no UV divergent terms appear within this formalism, one still has to introduce arbitrary parameters to regularize the otherwise ill-defined distributions. Regarding the Minkowski vacuum energy, one can see from Eq.~\eqref{Eq:DistFields} that the distributions are expanded in terms of normal-ordered free fields, which implies a vanishing vacuum energy density, regardless of the particular choice of renormalization of the distributions. The normal ordering here may be thought of as removing $\langle\rho\rangle$ by fiat. In light of its irrelevance to flat space calculations in QFT, and its apparent sensitivity to high-energy physics, we should not be surprised that a rigorous construction of QFT would consciously exclude vacuum energy.

\subsection{QFT on curved spacetimes}
\label{CurvedQFT}
Instead of altering the conceptual foundations of general relativity to fit particle physics, some physicists have instead attempted to formulate QFTs on a classical curved spacetime background. This provides a different ``first step'' to unifying the two disciplines. One advantage to this approach is that it is on much more sound mathematical footing than standard treatments of QFT. The clarity that comes with mathematical rigour helps for understanding the nature of assumptions that are needed for defining products of quantum fields. In particular, careful attention should be paid to the splitting procedures used for defining time-ordered products of operators. The downfall of such rigour, however, is that realistic interactions cannot yet be formulated fully as models of the axioms. A mix of methodology is therefore the clearest way forward.
    
As discussed in the previous section, point-splitting procedures have been successfully employed in the construction of quantum electrodynamics, and more local modifications are currently used for generalizing QFT to generically curved spacetimes. Many people are working on defining QFTs in curved spacetimes, but the most demanding requirements of locality come from the work of Hollands and Wald (\citeyear{HollandsWald2001,HollandsWald2002,HollandsWald2008,HollandsWald2010}). A key procedure in their construction of local, covariant time-ordered products is a modified version of the Epstein-Glaser point splitting prescription.

The Epstein Glaser approach to defining operator-valued distributions is more local than the standard momentum space cutoff approaches, in that it can be done in small neighbourhoods of coordinates in position space. Hollands and Wald note, however, that

\begin{quote}
    the Epstein-Glaser method is not local in a strong enough sense for our purposes, since we need to ensure that the renormalized time ordered products will be local, covariant fields. A key step in the Epstein-Glaser regularization procedure is the introduction of certain ``cutoff functions'' of compact support in the ``relative coordinates'' that equal 1 in a neighborhood of [conincident points\ldots These] will not depend only on the metric in an arbitrary small neighborhood of $p$ and, thus, will not depend locally and covariantly on the metric in the sense required by condition t1 [of locality and general covariance]. There does not appear to be any straightforward way of modifying the Epstein-Glaser regularization procedure so that the resulting extension [\ldots] will satisfy property t1. In particular, serious convergence difficulties arise if one attempts to shrink the support of the cutoff functions (Hollands and Wald 2002, p. 322).
\end{quote}

Since they aim to define quantum fields on generic globally hyperbolic spacetimes, Hollands and Wald aim to respect the restrictions imposed by the general covariance of general relativity, and therefore to define time-ordered products only in terms of local neighbourhoods of points in the spacetime. Their strategy is to use the equivalence principle to note that the neighbourhood of a point in a generically curved spacetime looks ``flat'' to leading order:

\begin{quote}
    Although it is true that the leading order divergences [\ldots] will be essentially the same as in flat spacetime, in general there will be sub-leading-order divergences that are sensitive to the presence of curvature and are different from the divergences occurring for the corresponding [condition] in flat spacetime. Nevertheless, we [show] that any local, covariant distribution that satisfies our scaling, smoothness, and analyticity conditions admits a ``scaling expansion'' about [coincident points]. This expansion expresses [\ldots] as a finite sum of terms plus a remainder term with the properties that (i) each term in the finite sum is a product of a curvature term times a distribution in the relative coordinates that corresponds to a Lorentz invariant distribution in Minkowski spacetime [\ldots] and (ii) the remainder term admits a unique, natural extension to the [coincident limit] (p. 323).
\end{quote}
This results in a specific form of the operator product expansion discussed above, where one first defines a short distant expansion of the c-number distribution, and uses that in the overall definition of the local covariant field operators. Due to the lack of symmetries in generically curved spacetimes, QFTs cannot generically rely on the concepts of large scale Lorentz covariance, a well-defined frequency splitting procedure, or a privileged, Lorentz-invariant vacuum state. In the generic case of QFT on a classical spacetime background, then, one must depend only on the highly local properties of the fields, defined in with respect to the spacetime metric in a generally covariant manner. In this case, since there is no globally defined Lorentz-invariant vacuum state, there is no issue of regularizing vacuum energy in the standard way. In a later essay, \textcite{HollandsWald2008} argue that a definition of QFTs in terms of the operator product expansion coefficients---when placed in appropriately symmetric spacetimes required to define a unique vacuum state---will have nonzero vacuum expectation values. They speculate that nonperturbative effects for interacting, non-Abelian QFTs may lead to vanishingly small residue terms in the stress-energy vacuum expectation value, which could explain the observed value of the cosmological constant \parencite{HollandsWald2008}. Given the current state of defining QFTs on curved spacetimes, however, vacuum expectation values play an unimportant role, and vacuum energy is only renormalized to first order, depending on a free parameter as in the case of dimensional regularization (cf. \cite[Eq. (9)]{HollandsWald2008}. Certainly, the concept of a globally well-defined, position-invariant vacuum energy density does not fit with this framework.

\section{Conclusions: Does QFT predict the value of the vacuum energy?}
\label{Conclusion}
Since vacuum energy is not fully renormalizable, the ``old-fashioned'' view of QFTs---as only well-defined if renormalizable---would lead one to believe that the vacuum expectation value of energy is an ill-defined concept in this framework.\footnote{Technically, old demands of renormalizability were imposed on the S-matrix of a model of QFT, believed to encode all physically meaningful content of scattering amplitudes and other dynamics \parencite{Dyson2,tHooftVeltman}. The QFTs comprising the standard model of particle physics are all renormalizable, despite the fact that the vacuum energy for each is nonrenormalizable. If one demands renormalizability of a model in terms of its S-matrix, additional nonrenormalizable structure that can be extracted from the action should be thought of as ill-defined surplus structure, about which the theory remains silent.} But with the interpretation of the standard model as an EFT, full renormalizability is no longer a strict requirement. Using a Euclidean lattice formulation of a particular model of QFT with a momentum regulator (cf. Section~\ref{EFT}), nonrenormalizable terms in the Lagrangian are suppressed by powers of the cutoff. If the cutoff is taken to be a physically meaningful quantity, then there is an accompanying physical interpretation that, at energy scales far below the cutoff, nonrenormalizable terms will be heavily suppressed and therefore of little relevance. These arguments are based on the renormalization group analysis of irrelevant terms in the Lagrangian; marginal terms are the ones found to play a role at all energy scales, while relevant terms grow in relative importance at low energies.

Unfortunately for the standard EFT view, the vacuum energy is one of two seemingly physically significant quantities in the standard model that are relevant terms under renormalization group flow.\footnote{The other, of course, being the Higgs mass. In that case the physical significance is undeniable, since the Higgs boson has been discovered, and has mass about 125GeV \parencite{CMS2019}. The physical significance of vacuum energy is a bit less direct, and is subject to criticism. Aside from the criticism raised in this paper, see \textcite{BianchiRovelli}.} The EFT approach licences taking nonrenormalizable terms to be physically significant, but vacuum energy does not fit into the standard physical interpretation, since it is not suppressed by powers of the cutoff. By insisting that the vacuum energy is physically significant, this problem of nonrenormalizability is one part of the cosmological constant problem. In response, one can reject the assumption that the vacuum energy as predicted by the standard model is physically meaningful, or one can weaken the demand of renormalizability to understand what QFTs tell us about the value of the vacuum energy.

I have adopted this latter approach in this paper. By dropping the requirement of renormalizability, we are left with either regularized, or one-loop renormalized quantities describing vacuum energy density. In the Introduction, I claimed that two minimal Reasonable Requirements for a quantity to count as a candidate prediction are the following.\\

\textbf{Minimal Requirements for Candidate Predictions:} 
\textit{In order for a quantity within a model of QFT to count as a candidate prediction of some corresponding physical quantity, it must be the case that: (1) the quantity is largely insensitive to the regularization procedure; and (2) it is largely insensitive to changes to the value of the regulator.}\\

Since regularization procedures in QFT are somewhat arbitrary, and usually the regulator disappears from the final prediction of a physical quantity, one might expect that full independence of the regularization technique be required. This seems like too strict a condition, however, when one considers that regularization changes the form of a model of QFT. Different changes will lead to different regulators, and full renormalization is required to make these different approaches agree. Under the standard EFT view, one can think of the different regularization schemes as different ways of parameterizing our ignorance of high-energy physics. One can only trust the predictions of an EFT when these differences wash out, which happens when the Minimal Requirements are satisfied.

For the orthodox regularization schemes discussed in Section~\ref{Orthodox}, a purely regularized vacuum energy density fails to meet either of the Minimal Requirements. The lattice regularized expression depends on the large-momentum cutoff $\mu$ as $\langle\rho\rangle \sim \mu^4$, while the dimensionally regularized term depends on the small deviation from four dimensions $\epsilon$ as $1/\epsilon$. Small changes to these regulators will lead to large changes in $\langle\rho\rangle$. Further, the expressions in Eqs.~\eqref{cutoffrho} and~\eqref{rhodimreg} are quite different, so the value of $\langle\rho\rangle$ is sensitive to the regularization procedure. The two vacua described under these procedures even differ in their equation of state.

If one rejects requirement (1), and takes the one-loop renormalized value of $\langle\rho_{SM}\rangle$ as a first order prediction, then one has a candidate prediction for vacuum energy density that can be used to motivate a cosmological constant problem. However, there are two issues here. First, renormalized quantities in QFTs aren't taken as predictions of some physical quantity. After renormalization, the physical value is measured from experiment and input into the theory. In this sense, Eq.~\eqref{SMrho} would not count as a prediction of vacuum energy density, but would be tuned to give the measured value. The instability of $\langle\rho\rangle$ under radiative corrections makes this tuning impossible perturbatively; so the failure of naturalness prevents a consistent tuning. Second, this prediction is not straightforwardly compatible with EFT, which I have taken to justify the search for a nonrenormalizable candidate prediction of $\langle\rho\rangle$.

To see this, consider the case of the Fermi model of weak interactions. This is a model in which four fermions---a proton, neutron, electron, and muon---all interact at a point. This model is not fully renormalizable, but it is one-loop renormalizable. Physicists used this model to make predictions at the one-loop level, even though higher order terms were known to diverge. The success of the Fermi model can be explained by noting that it is an effective theory of the electroweak model. Nonrenormalizable terms that appear above the one-loop level are due to the absence in the Fermi model of the W boson to mediate the four-fermion interaction. These divergent terms end up being irrelevant under renormalization group flow, so the mass scale ($M_W\approx 80$GeV) of the W boson in an effective modification of the Fermi theory suppresses the divergent terms. Successful use of Fermi theory for low energy ($m_F\approx 10$MeV) predictions is justified by the EFT framework, since $m_F \ll M_W$.

When looking at the standard model as an EFT, one might hope that a similar story can be told for the vacuum energy density. In some successor theory, the relevant energy scale there will suppress the extremely large value $\langle\rho_{SM}\rangle$. This is one way of expressing the requirement that vacuum energy be natural. However, $\langle\rho\rangle$ is \textit{relevant} under renormalization group flow, and should depend quartically on a cutoff supplied by a theory to which the standard model is effective. Given that the quantity $\langle\rho\rangle$ is so sensitive to the value of the regulator beyond one-loop, I take this to disqualify it as a candidate prediction. From within the standard model, we have reason to believe that $\langle\rho\rangle$ depends sensitively on the details of high-energy physics, and therefore falls outside the scope of EFT. Even if one rejects the Minimal Requirements and takes $\langle\rho_{SM}\rangle$ as a candidate prediction, when factoring in all fundamental fields in the standard model, the value $\langle\rho_{SM}\rangle$ is approximately 55 orders of magnitude too large. While much smaller than the often quotes 120 orders of magnitude, this is still a remarkably bad prediction. Given its independence from all predictions within orthodox QFT, one should therefore be skeptical of such a prediction (cf. \textcite{KoberinskiSmeenkCCP} for further discussion).

If standard EFT methods do not provide a candidate prediction of $\langle\rho\rangle$, should we expect more rigorous extensions of QFT to incorporate vacuum energy? Normal ordering procedures---including the Epstein-Glaser approach---define all vacuum expectation values to vanish, so in a sense these approaches ``renormalize'' the vacuum energy density to zero. Normal ordering is typically defined for free fields, and as we have seen for orthodox approaches, the presence of interactions can spoil renormalizability. The Epstein-Glaser point splitting approach treats regularization and renormalization in a very different way, and relates UV divergences to ill-defined products of distributions at singular points. By carefully splitting distributions, one avoids divergent integrals. However, there is still freedom in the definition of these distributions, and this amounts to renormalization in a similar manner: free parameters in the theory must be fixed by experiment. These numerical distributions are then used to define operator-valued distributions, which include normal-ordered free fields. So in this formalism, normal ordering is directly connected to meaningful time-ordered products (equivalently, n-point functions), and so Epstein-Glaser point splitting leads to a vanishing vacuum expectation value of all quantities, energy density included. 

Finally, the Hollands and Wald approach to QFTs in curved spacetime significantly alters and extends the core concepts of perturbative QFT on Minkowski spacetime. Their approach to merging QFT and general relativity is to reformulate the principles of QFT to be compatible with the spacetime structure of generic globally hyperbolic solutions to the Einstein field equations. For QFTs on curved spacetimes, analogs to Lorentz covariance and global frequency splitting---general covariance and the microlocal spectrum condition---change the mathematical formalism significantly. Even more significantly, vacuum states are generically ill-defined, and so vacuum expectation values cannot be the primary building blocks of n-point functions. \textcite{HollandsWald2008} have suggested that the operator product expansion coefficients could be used to define a model of QFT. In highly symmetric cases, one may recover a vacuum state as a derived concept; it would then make sense to discuss vacuum energy densities, but this would be highly dependent on the particular spacetime chosen. A Lorentz-invariant vacuum energy density is not a generic feature of local covariant QFT, and there is no guarantee that the Minkowski prediction in this radically different formalism would agree with one of the orthodox schemes. These extensions of the standard QFT framework support the conclusion that QFTs (considered as EFTs or otherwise) do not properly include vacuum energy density.

\subsection{Verdict: No \texorpdfstring{$\langle\rho\rangle$}{<rho>} from the standard model}
Does QFT in general---or the standard model in particular---predict a vacuum expectation value of energy density? According to the Minimal Requirements motivated by viewing the standard model as an EFT, it does not. We have seen that under the orthodox approaches to regularization, vacuum energy density varies significantly with the choice of regularization scheme---lattice regularization or dimensional regularization---and the ``predicted'' value of $\langle\rho\rangle$ is sensitively dependent on the value of the regulator. If we reject Requirement (1), then one might be in a position to pick the dimensionally regularized quantity as a candidate prediction. In order to do so, one must first acknowledge that $\langle\rho\rangle$ falls outside the domain of typical quantities in EFTs. One of the remarkable features of thinking of the standard model as an EFT is that ``the details of physics below the cutoff have \textit{almost no empirical consequences} for large-scale physics''  \parencite[p. 10, emphasis original]{WallaceQFT}. By rejecting Requirement (1), we are admitting that, for some physically meaningful quantities in the EFT, the choice of regularization scheme---of how to parameterize ignorance of high-energy physics---makes a considerable difference to the predicted value of that quantity within QFT. Moreover, this would also amount to claiming that dimensional regularization is \textit{the correct} way to do so in this instance. Instead, one should acknowledge that the sensitivity to regularization scheme is a sign that the quantity falls outside the scope of the EFT.

If one still insists on prioritizing dimensional regularization, then one must renormalize the vacuum energy density at one-loop in order to satisfy Requirement (2). Though the value $\langle\rho_{SM}\rangle = -2 \times 10^8 GeV^4$ appears insensitive to the regulator (Requirement (2)), this is only because high sensitivities at higher orders are hidden by brute truncation. The quantity is not perturbatively renormalizable, and new sensitivities to the regulator $\epsilon$ will appear at each order. Further, there is no principled reason to pick any given order at which to renormalize. Since the divergences are of the same character at each order, and since the regulator makes the same order of contributions at each order, the only principled choice is to renormalize nonperturbatively. Since this cannot be done with the vacuum energy density, there is no reason to renormalize perturbatively at any particular order. If renormalization at, e.g, one-loop level yielded a sensible prediction, then there might be a post-hoc justification. But since $\langle\rho_{SM}\rangle$ is still so far off the from the observed value, this seems like an unjustified relaxation of the Minimal Requirements, and indicates that the quantity $\langle\rho_{SM}\rangle$ lacks physical significance.

I argue that \textit{both} Minimal Requirements are needed for a quantity to count as a candidate prediction of some corresponding physical quantity under the EFT framework. This is a hallmark of all other predictions of QFTs, and is \textit{not} satisfied in the case of vacuum energy density. Since there is no direct evidence necessitating a physically significant vacuum energy density in QFTs, I do not think we have grounds for a candidate prediction.\footnote{Cf. \textcite{KoberinskiCCP} for an argument that the Casimir effect and Lamb shift do not license the inference to a constant vacuum expectation value of energy.} 
Under the standard view, vacuum energy density should be treated as analogous to droplet formation in fluid mechanics: outside the scope of the EFT, and requiring the details of the high-energy theory in order to make sense. Just as we don't expect fluid mechanics to provide the details of droplet formation, we should not expect the standard model to predict the value of vacuum energy density. To be clear, I have not argued that the concept of vacuum energy density is meaningless; it is simply outside the scope of EFT. An alternative approach is to extend and modify QFT to better fit with the principles of general relativity, as outlined in Sec.~\ref{CurvedQFT}. In particular, the concept of the vacuum will likely require significant revision. The absence of $\langle\rho\rangle$ from local extensions of QFT mentioned in Sec.~\ref{PointSplit} suggests further that vacuum energy is not a proper part of the physical content of QFT. The cosmological constant problem should be understood as indicating some inconsistency in merging Minkowski QFTs with general relativity at the level of EFTs \parencite{KoberinskiSmeenkCCP}. In particular, the presence of a large effective cosmological constant undermines the initial assumption that Minkowski spacetime is a good approximation to the more realistic curved spacetime. The work of Hollands and Wald highlights how much of the formalism may need to change if one wants to make QFTs conceptually compatible with the general covariance and locality of general relativity. Perhaps the resulting conceptual clarity will also serve to clear up the concept of vacuum energy density as well.

The cosmological constant problem does require some sort of (dis)solution. By investigating the foundations of QFT, it is increasingly clear that at least part of the problem lies in accepting that the standard model provides a candidate prediction of $\langle\rho\rangle$.

\section*{Acknowledgements}
Removed for review
I am grateful to Chris Smeenk, Robert Brandenberger, Doreen Fraser, and the UCI Philosophy of Physics Research group for helpful feedback on early drafts of this paper, as well as the comments from two anonymous reviewers. This work was supported by the Social Sciences and Humanities Research Council of Canada, and the John Templeton Foundation Grant 61048, \textit{New Directions in Philosophy of Cosmology}. The opinions expressed in this publication are those of the author and do not necessarily reflect the views of the John Templeton Foundation.
\pagebreak
\printbibliography

\end{document}